\documentclass[12pt]{article}

\usepackage{amsmath,amssymb}
\usepackage{graphicx,a4,psfrag,here,epsfig,epsf}
\usepackage{pstricks}
\usepackage{pst-node}
\usepackage{citesort}
\allowdisplaybreaks[1]

\begin{document}
\newcommand{\sh}{\not\!}
\newcommand{\lag}{{\mathcal L}}
\newcommand{\ol}{\overline}
\newcommand{\co}{\; \; ,}
\newcommand{\per}{ \; .}
\newcommand{\nn}{\nonumber}
\newcommand{\tr}[1]{\langle #1 \rangle}

\newcommand{\lt}{\left}
\newcommand{\rt}{\right}
\renewcommand{\d}{\partial}
\newcommand{\fc}{\frac}

\newcommand{\rR}{R}
\newcommand{\rr}{{\cal R}}
\newcommand{\mc}{\multicolumn{2}{|c|}}
\newcommand{\mcc}{\multicolumn{3}{|c|}}
\newcommand{\nl}{\!\! & \!\!}
\newcommand{\av}[1]{\langle {#1}\rangle}

\def\beq{\begin{equation}}
\def\eeq{\end{equation}}
\def\bea{\begin{eqnarray}}
\def\eea{\end{eqnarray}}
\def\eq{\begin{eqnarray}}
\def\en{\end{eqnarray}}
\newcommand{\bes}{\begin{split}}
\newcommand{\ees}{\end{split}}
\newcommand{\bed}{\begin{displaymath}}
\newcommand{\eed}{\end{displaymath}}
\def\prop{{\mathcal D}}

\newcommand{\mpi}{M_\pi^2}
\newcommand{\mk}{M_K^2}
\newcommand{\me}{M_\eta^2}
\newcommand{\equ}{\,=\,}
\newcommand{\no}{\nonumber}
\newcommand{\noh}{\nonumber \hskip -10cm}
\newcommand{\Order}{\mathcal{O}}
\newcommand{\Lagr}{\mathcal{L}}
\newcommand{\M}{\mathcal{M}}
\numberwithin{equation}{section}
\newcommand{\ed}{\end{document}}
\newcommand{\scs}{\, , \,}
\newcommand{\sem}{\, ; \,}
\newcommand{\nnnl}{\nonumber\\}
\newcommand{\fs}{\, . \,}
\newcommand{\be}{\begin{eqnarray}}
\newcommand{\ee}{\end{eqnarray}}
\newcommand{\bra}{\right\rangle}
\newcommand{\bla}{\left\langle}
\newcommand{\word}[1]{{\mbox{{#1}\,}}}
\newcommand{\words}[1]{{\mbox{\small{#1}}}}
\newcommand{\wordt}[1]{{\mbox{\tiny{#1}}}}

\def\query#1{\marginpar{\begin{flushleft}\footnotesize#1\end{flushleft}}}%

\begin{titlepage}

\vspace{1cm}
\begin{center}{\Large\bf Generating functional of ChPT at one loop for non-minimal operators}

\vspace{0.5cm}
\today

\vspace{0.5cm}
Andria Agadjanov$^{a,b}$, Dimitri Agadjanov$^{a,b}$,
Anzor Khelashvili$^{b,c}$ and\\ Akaki  Rusetsky$^a$

\vspace{2em}
\footnotesize{\begin{tabular}{c}
$^a\,$ Helmholtz-Institut f\"ur Strahlen- und Kernphysik (Theorie) and\\
Bethe Center for Theoretical Physics,\\
 $\hspace{2mm}$   Universit\"at Bonn,
Nussallee 14-16, D-53115 Bonn, Germany\\[2mm]
$^b\,$ St. Andrew the First-Called Georgian University of the Patriarchate of Georgia,\\ Chavchavadze Ave. 53a, 0162, Tbilisi, Georgia\\[2mm]
$^c\,$ Institute of High Energy Physics, Ivane Javakhishvili Tbilisi State University,\\ University St. 9, 109, Tbilisi, Georgia
\end{tabular}  }

\vspace{1cm}

\begin{abstract}
The divergent part of the one-loop effective action in Chiral Perturbation
Theory with virtual photons has been evaluated in an arbitrary covariant gauge.
The differential operator, that emerges in the functional determinant, is
of a non-minimal type, for which the standard
heat kernel methods are not directly applicable. Both $SU(2)$ and $SU(3)$ cases have
been worked out. A comparison with existing results in the literature
is given.
\end{abstract}

\vspace{1cm}
\footnotesize{\begin{tabular}{ll}
{\bf{Pacs:}}$\!\!\!\!$& 11.10.Gh, 11.30.Rd, 12.39.Fe, 13.40.Ks
\\
{\bf{Keywords:}}$\!\!\!\!$& Chiral symmetries, chiral Lagrangians,
electromagnetic interactions,\\
& ultraviolet divergences
\\
\end{tabular}}
\end{center}
\end{titlepage}

\setcounter{page}{2}


\section{Introduction}

Isospin-breaking corrections to the hadronic observables emerge from two
intrinsically distinct sources. The so-called ``strong'' isospin breaking,
which is due to the difference of the $u$- and $d$-quark masses,
is embedded in the
standard framework of Chiral Perturbation Theory (ChPT) from the outset~\cite{Weinberg,GL}.
The electromagnetic isospin breaking, caused by a presence of
the virtual photons, can be systematically included by the use of the spurion
technique~\cite{Urech} (see also~\cite{Rupertsberger,MMS,Knecht-Urech}).
ChPT with photons and leptons, which is the
low-energy effective theory of the Standard Model, can be also
constructed~\cite{Neufeld}. The method has been extended to include the
baryon sector of ChPT as well~\cite{Meissner-baryon}.
Last but not the least, including virtual photons becomes inevitable, if
an attempt is made to
describe metastable bound states of hadrons -- the
 so-called hadronic atoms -- which are kept together predominately by the
Coulomb force and decay mainly through strong interactions (for a recent
review, see, e.g.~\cite{physrep} and references therein).
Note, however, that defining the splitting
of the hadronic observables into the ``purely strong'' and ``electromagnetic''
contributions in ChPT is, in general, an ambiguous procedure due to the
renormalization group running of the parameters of the underlying
theory~\cite{Scimemi,CD} (for related work on the subject, see
also~\cite{Moussallam,Anant}).

One of the issues, which should be addressed during the construction of the
low-energy effective theory of the Standard Model, is the gauge dependence
of the parameters of the effective Lagrangian (more precisely, the dependence of these
parameters on the gauge fixing of the electromagnetic field). It should be
pointed out that, in the matching of the effective theory to
the Standard Model,
non-perturbative QCD effects are essentially involved, and the study
of the gauge dependence
can provide valuable information about the relation of the couplings
of the effective Lagrangian to the fundamental parameters of the underlying
theory. We also expect that such
investigation might
be helpful for the interpretation of the lattice QCD results including
electromagnetic effects, which started to appear recently
(see, e.g.,~\cite{Basak,BMW,Blum,Ukita,Fodor}).

The gauge-dependence of certain $O(e^2p^2)$ and
$O(e^4)$ electromagnetic
low-energy constants (LECs) in ChPT has been investigated in the
past (see, e.g.,~\cite{Scimemi,Moussallam,Anant}) but a systematic analysis
of the problem is still missing. One obvious
reason for this is that the one-loop generating functional in
ChPT up to now has been calculated only in the Feynman gauge. In other gauges,
the differential operator, which emerges in the functional determinant, is
of a non-minimal type. Standard heat kernel methods\footnote{See, 
e.g., Ref.~\cite{Vassilevich} for a review. For the extension of the method to the case when fermions are present, see Refs.~\cite{Berezinian}} are not applicable
for such kernels. In the literature, one finds
examples of the calculation of the second
Seeley-DeWitt coefficient for the non-minimal operators, which determines
the UV-divergent part of the effective action at one
loop~\cite{Avramidi,GK,Gusynin,Vilkovisky,Stepanyantz,Shore}, see also
Ref.~\cite{Anant_review} for a review and comparison of different
methods. However, to the best of our knowledge, the general methods so far
have not been applied to the particular problem we are interested in.
The aim of the study, carried out in this brief note, is to close the gap and to
present a calculation of the divergent
part of the one-loop effective functional in ChPT with virtual photons
in an arbitrary covariant gauge.

The layout of the paper is as follows. In the
section~\ref{sec:notations} we collect all the relevant notation and display the $O(p^4)$
Lagrangian in the 3-flavor ChPT. In section~\ref{sec:divergent} we
write down the one-loop effective action with virtual photons. 
The calculation of the divergent part of the functional determinant in the
arbitrary covariant gauge is presented in section~\ref{sec:a.neq.1} and
the renormalization is carried out in section~\ref{sec:LECs}, where the
divergent part of the LECs both in 3-flavor and 2-flavor cases are displayed.
In the section~\ref{sec:comparison},
comparison to the results, available in the literature, has been carried
out. Finally, section~\ref{sec:concl} contains a short summary of our findings.

\section{ChPT with virtual photons}
\label{sec:notations}

This section collects the notations, which will be used in the following.
The lowest order Lagrangian of ChPT with virtual photons in case of three
flavors is given by
\eq\label{eq:L2}
{\cal L}_2=\frac{F_0^2}{4}\langle d_\mu U d^\mu U^\dagger
+\chi U^\dagger + U\chi^\dagger\rangle+C\langle UQU^\dagger Q\rangle,
\en
where $U$ is a unitary $3\times 3$ matrix containing eight Goldstone boson fields,
the brackets stand for the traces in flavor space, and
\eq
d_\mu U&=&\partial_\mu U-iR_\mu U+iUL_\mu,
\\[2mm]
R_\mu&=&v_\mu + a_\mu+ A_\mu Q ,\quad\quad  L_\mu=v_\mu - a_\mu+ A_\mu Q\, ,
\\[2mm]
\chi&=&2B_0(s+ip)\, .
\en
Here $Q=e\,\mbox{diag}\,(2/3,-1/3,-1/3)$ is the charge matrix of the quarks,
$s,p,v_\mu,a_\mu$ are the external scalar, pseudoscalar, vector and axial-vector
sources, respectively, and $A_\mu$ is the electromagnetic field. The
$O(p^2)$ low-energy constant $F_0$ is the pion decay constant
in the chiral limit, $B_0$ is related to the quark condensate and
$C=F_0^4Z$ describes the $O(e^2)$ electromagnetic mass splittings
of Goldstone bosons:
\eq
M_{\pi^+}^2-M_{\pi^0}^2=M_{K^+}^2-M_{K^0}^2=2e^2F_0^2 Z\, .
\en
The Lagrangian~(\ref{eq:L2}) should be supplemented by the electromagnetic field Lagrangian
\eq
L_\gamma=-\frac{1}{4}\,F_{\mu\nu}F^{\mu\nu}
-\frac{1}{2a}\,(\partial^\mu A_\mu)^2\, ,
\en
where $a$ denotes the gauge fixing parameter.

The full Lagrangian at next-to-leading order is given in Ref.~\cite{Urech}.
It contains strong and electromagnetic terms
\eq\label{Lem}
{\cal L}_4 &=&
L_1\langle d^\mu U^\dagger d_\mu U\rangle^2
+L_2\langle d^\mu U^\dagger d^\nu U\rangle \langle d_\mu U^\dagger d_\nu U\rangle
\nn\\[2mm]
&+&L_3\langle d^\mu U^\dagger d_\mu U d^\nu U^\dagger d_\nu U\rangle
+L_4\langle d^\mu U^\dagger d_\mu U\rangle  \langle \chi^\dagger U+\chi U^\dagger\rangle
 \nn\\[2mm]
&+&L_5\langle d^\mu U^\dagger d_\mu U\,(\chi^\dagger U+U^\dagger\chi)\rangle
+L_6\langle \chi^\dagger U+U^\dagger\chi\rangle^2
+L_7\langle \chi^\dagger U-U^\dagger\chi\rangle^2
 \nn\\[2mm]
&+&L_8\langle \chi^\dagger U\chi^\dagger U+\chi U^\dagger\chi U^\dagger\rangle
-iL_9\langle d^\mu U d^\nu U^\dagger R_{\mu\nu}+d^\mu U^\dagger d^\nu U L_{\mu\nu}\rangle
 \nn\\[2mm]
&+&L_{10}\langle R^{\mu\nu}UL_{\mu\nu}U^\dagger\rangle
+H_1\langle R^{\mu\nu}R_{\mu\nu}+L^{\mu\nu}L_{\mu\nu}\rangle
+H_2\langle  \chi^\dagger\chi\rangle
\nn\\[2mm]
&+&F^2_0 \big\{ K_1\langle d^{\mu}U^\dagger d_{\mu} U\rangle
\langle Q^2 \rangle +K_2 \langle d^{\mu}U^\dagger d_{\mu}U\rangle\langle
QU QU^\dagger \rangle
\nn\\[2mm]
&+& K_3\big(\langle d^{\mu}U^\dagger Q U \rangle\langle d_{\mu}U^\dagger Q U
\rangle + \langle d^{\mu}U Q U^\dagger \rangle\langle
d_{\mu}U Q U^\dagger\rangle\big)
\nn\\[2mm]
&+& K_4\langle d^{\mu}U^\dagger Q U\rangle\langle d_{\mu}U Q U^\dagger \rangle
+K_5\langle \big(d^{\mu}U^\dagger d_{\mu}U+d^{\mu}U d_{\mu}U^\dagger\big)Q^2\rangle
\nn\\[2mm]
&+&K_6\langle d^{\mu}U^\dagger d_{\mu}UQU^\dagger QU + d^{\mu}U
d_{\mu}U^\dagger QUQU^\dagger\rangle
\nn\\[2mm]
&+&K_7\langle\chi^\dagger U+U^\dagger \chi\rangle\langle Q^2\rangle+ K_8\langle\chi^\dagger U+U^\dagger\chi\rangle\langle Q U Q U^\dagger \rangle
\nn\\[2mm]
&+& K_9\langle(\chi U^\dagger + U\chi^\dagger + \chi^\dagger U+U^\dagger\chi )Q^2\rangle
\nn\\[2mm]
&+&K_{10}\langle(\chi U^\dagger + U\chi^\dagger )QUQU^\dagger
+(\chi^\dagger U+U^\dagger\chi )Q U^\dagger Q U\rangle
\nn\\[2mm]
&+& K_{11}\langle(\chi U^\dagger - U\chi^\dagger )QUQU^\dagger
+(\chi^\dagger U-U^\dagger\chi )Q U^\dagger Q U\rangle
\nn\\[2mm]
&+& K_{12} \langle d_{\mu}U^\dagger [c_{ R}^{\mu}Q,Q]U
+d_{\mu}U[c_{ L}^{\mu}Q,Q]U^\dagger\rangle
\nn\\[2mm]
&+& K_{13} \langle c_{ R}^{\mu}Q U c_{ L \mu}Q U^\dagger\rangle +
K_{14}\langle c_{ R}^\mu Q c_{ R \mu} Q + c_{ L}^\mu Q
c_{ L\mu}Q \rangle\big\}
\nn\\[2mm]
&+&F^4_0\big\{K_{15}\langle QUQU^\dagger\rangle^2 + K_{16}\langle QUQU^\dagger\rangle\langle Q^2\rangle + K_{17}\langle Q^2\rangle^2 \big\},
\en
where
\eq
c_I^\mu Q&=&-i[I^\mu,Q]\, ,\quad\quad I=L,R\, ,\\[2mm]
R_{\mu\nu}&=&\partial_\mu R_\nu-\partial_\nu R_\mu-i[R_\mu,R_\nu]\, ,\quad
L_{\mu\nu}=\partial_\mu L_\nu-\partial_\nu L_\mu-i[L_\mu,L_\nu]\,.
\en
The coefficients $L_i,H_i,K_i$ cancel the UV divergences, arising from the divergent
part of the one-loop effective action
\eq\label{eq:LHK}
L_i&=&\Gamma_i\lambda+L_i^r(\mu)\, ,\nn\\[2mm]
H_i&=&\Delta_i\lambda+H_i^r(\mu)\, ,\nn\\[2mm]
K_i&=&\Sigma_i\lambda+K_i^r(\mu)\, ,\nn\\[2mm]
\lambda&=&\frac{\mu^{d-4}}{16\pi^2}\,\biggl\{\frac{1}{d-4}-\frac{1}{2}\,
(\Gamma'(1)+\ln 4\pi+1)\biggr\}\, ,
\en
where $d$ is the number of space-time dimensions and $\mu$ denotes the
scale of dimensional regularization. The aim of the present work is to determine the gauge-dependent part of $\Sigma_i$. The quantities $\Gamma_i,\Delta_i$, as well as $\Sigma_i$ at $a=1$ are already available in the literature.

\section{One-loop effective action}
\label{sec:divergent}

The one-loop generating functional for the connected Green functions is given by
\eq
e^{iZ(v,a,s,p)}=\int dU dA_\mu\,e^{i\int dx ({\cal L}_2+{\cal L}_4)}\, ,
\en
where the integral is evaluated in the semi-classical approximation.
To this end, we expand the
fields $U(x), A_\mu(x)$ around the solutions of the classical
equations of motion $\bar U,\, \bar A_\mu$:
\eq\label{eq:generating}
U&=&ue^{i\xi/F_0}u=u\left({\bf 1}+i\frac{\xi}{F_0}
-\frac{1}{2}\frac{\xi^{2}}{F^2_0}+\cdots\right)u \no\\[2mm]
&=&\bar{U}+\frac{i}{F_0}u\xi u-\frac{1}{2F^2_0}u\xi^{2}u+\ldots \no\\[2mm]
A_{\mu}&=&\bar{A}_{\mu}+\epsilon_{\mu},
\en
where $\bar U=u^2$ and $\xi$ is a traceless hermitian matrix,
$\xi=\sum_{a}\xi^a\lambda^a$ (here, $\lambda^a$ denote the Gell-Mann matrices).

Next, we substitute this expansion in the action functional in
Eq.~(\ref{eq:generating}) and retain those terms, stemming from ${\cal L}_2$,
which are at most quadratic in $\xi,\epsilon_\mu$ (at this order, the fields in
${\cal L}_4$ can be replaced by the classical solutions). The calculations are conveniently done in Eucledean space. The Euclidean action functional then
becomes
\eq
S_{\sf E}=\int dx({\cal L}_2+{\cal L}_4)=
\int dx(\bar{\cal L}_2+\bar{\cal L}_4)
+ \frac{1}{2}\int dx\,\eta_A D^{AB}\eta_B,  ,
\en
where the Lagrangians $\bar{\cal L}_2,\bar{\cal L}_4$ 
are obtained from
${\cal L}_2,{\cal L}_4$ after continuation to Euclidean space and
 the substitution $U,A_\mu\to \bar U,\bar A_\mu$.
The fluctuations are collected in a single vector
\eq
\eta_A=(\xi_a,\epsilon_\mu) = (\xi_1,\ldots,\xi_8,\epsilon_0,\ldots,\epsilon_3)\, .
\en
The differential operator $D$ is defined as:
\eq
\label{det}
D&=&D_0+\omega,\nn\\[3mm]
D_0&=&\left( \begin{array}{cc}
-\Box\delta^{ab}&0\\
0&-\Box \delta^{\sigma\rho}+\left(1-\frac{1}{a}\right)\partial^\sigma\partial^\rho
\end{array}\right),\nn\\[3mm]
\omega&=&-\{Y_\mu, \partial_\mu\}-Y_\mu Y_\mu+\Lambda\, ,
\en
with
\begin{eqnarray}
Y_\mu=\left( \begin{array}{cc}
\Gamma^{ab}_\mu&X^{a\rho}_\mu\\
X^{\sigma b}_\mu&0
\end{array}\right),\quad\quad
\Lambda=\left( \begin{array}{cc}
\sigma^{ab}&\frac{1}{2}\gamma^{a\rho}\\
\frac{1}{2}\gamma^{\sigma b}&\rho \delta^{\sigma\rho}
\end{array}\right).
\end{eqnarray}
The elements of these matrices are given by the expressions:
\begin{eqnarray}
\label{YLambda}
\Gamma^{ab}_\mu&=&-\frac{1}{2}\tr{[\lambda^a,\lambda^b]\Gamma_\mu},\nn\\[2mm]
X^{a\rho}_\mu&=&-X^{\rho a}_\mu=X^a\delta^\rho_\mu,\quad X^a=-\frac{F_0}{4}\tr{H_L\lambda^a},\nn\\[2mm]
\sigma^{ab}&=&-\frac{1}{2}\tr{[\Delta_\mu,\lambda^a][\Delta_\mu,\lambda^b]} + \frac{1}{4}\tr{\{\lambda^a,\lambda^b\}\sigma}-\frac{F^2_0}{4}\tr{H_L\lambda^a}\tr{H_L\lambda^b}\nn\\[2mm]
&&-\frac{C}{8F^2_0}\left\{\tr{[ H_R+H_L,\lambda^a][ H_R-H_L,\lambda^b]+a\leftrightarrow b}\right\},\nn\\[2mm]
\gamma^{a\rho}&=&\gamma^{\rho a} = F_0\tr{\left([ H_R,\Delta^\rho]+\frac{1}{2}D^\rho H_L\right)\lambda^a},\nn\\[2mm]
\rho&=&\frac{3}{8}F^2_0\tr{H_L^2},
\end{eqnarray}
where
\begin{eqnarray}
D_\mu H_L&=&\partial_\mu H_L+[\Gamma_\mu, H_L]\nn\\[2mm]
\Gamma_\mu&=&\frac{1}{2}[u^\dagger,\partial_\mu u] - \frac{1}{2}iu^\dagger \bar R_\mu u - \frac{1}{2}iu \bar L_\mu u^\dagger,\nn\\[2mm]
 \bar R_\mu&=&v_\mu+a_\mu+Q\bar A_\mu\, ,\quad \bar L_\mu=v_\mu-a_\mu+Q\bar A_\mu\, ,
\nn\\[2mm]
\Delta_\mu&=&\frac{1}{2}u^\dagger d_\mu\bar U u^\dagger=-\frac{1}{2}ud_\mu\bar
U^\dagger u,\nn\\[2mm]
 H_R&=&u^\dagger Q u+uQ u^\dagger,\nn\\[2mm]
 H_L&=&u^\dagger Q u-uQ u^\dagger,\nn\\[2mm]
\sigma&=&\frac{1}{2}(u^\dagger\chi u^\dagger + u\chi^\dagger u).
\label{sada}
\end{eqnarray}
Thus, the Euclidean generating functional at one loop is given by
\eq
\label{Zp4}
Z(v,a,s,p)=\int dx(\bar{\cal L}_2+\bar{\cal L}_4)+ \frac{1}{2}\ln \det D,
\en
where all quantities are to be evaluated at the classical solutions $\bar U(x), \bar A_\mu(x)$. The determinant of the operator $D$ requires
renormalization, since it contains divergences of one-loop graphs.
As mentioned before, these divergences should be absorbed by the counterterms,
contained in $\bar{\cal L}_4$, see Eq.~(\ref{eq:LHK}).

Note that, if $a=1$, the expression in Eq.~(\ref{det}) turns into the
standard expression (see, e.g.,~\cite{Urech,Knecht-Urech}). The differential operator,
emerging there, is of a minimal type. The UV-divergent part thereof
can be found in a straightforward manner by using the well-known expression
\eq\label{eq:a=1}
\frac{1}{2}\,\ln\det D\biggr|_{a=1}
=-\lambda \int dx\, \mbox{tr}\,\left(\frac{1}{12}Y_{\mu\nu}Y_{\mu\nu}+\frac{1}{2}\Lambda^2\right) + \mbox{UV-finite part},
\label{Zmin}
\en
where ``tr'' means the trace in the multi-index $A=(a,\mu)$ and
\eq
Y_{\mu\nu}=\partial_\mu Y_\nu - \partial_\nu Y_\mu + [Y_\mu,Y_\nu]\, .
\en

\section{The case $a\neq 1$}
\label{sec:a.neq.1}

In general, when $a$ is not equal to 1, the differential operator 
in Eq.~(\ref{det}) is
of a non-minimal type and one has to resort to a different method for calculating the determinant.
In particular, in analogy to Ref.~\cite{Schweizer}, we
evaluate the UV-divergent part of the determinant by means of a straightforward expansion
in powers of $\omega$, see Eq.~(\ref{det})
\eq
\frac{1}{2}\ln \det (D_0+\omega)&=&\frac{1}{2}\ln \det D_0
+\frac{1}{2}\mbox{Tr}(\emph{D}^{-1}_0\omega)\nn\\[2mm]
&-&\frac{1}{4}\mbox{Tr}(\emph{D}^{-1}_0\omega\emph{D}^{-1}_0\omega)+
\frac{1}{6}\mbox{Tr}(\emph{D}^{-1}_0\omega\emph{D}^{-1}_0\omega\emph{D}^{-1}_0\omega)\nn\\[2mm]
&-&\frac{1}{8}\mbox{Tr}(\emph{D}^{-1}_0\omega\emph{D}^{-1}_0\omega\emph{D}^{-1}_0\omega\emph{D}^{-1}_0\omega)+\mbox{UV-finite part}\, ,
\label{Zexp}
\en
where the symbol ``Tr'' stands for the trace in both coordinate and
matrix indices. It is easy to truncate this
expression so as to retain only those terms that contain UV divergences.
If $a=1$, the divergent part of the expression above reproduces the already
known result, see Eq.~(\ref{eq:a=1}).

The matrix elements of the operator $\emph{D}^{-1}_0$ have the following form:
\eq
\tr{x|(\emph{D}^{-1}_0)^{ab}|y}&=&-\delta^{ab}\Delta(x-y),\nn\\[2mm]
\tr{x|(\emph{D}^{-1}_0)^{\rho\sigma}|y}&=&-\delta^{\rho\sigma}\Delta(x-y)
+\Delta^{\rho\sigma}(x-y),\nn\\[2mm]
\tr{x|(\emph{D}^{-1}_0)^{\rho b}|y}&=&0\, ,\quad\quad
\tr{x|(\emph{D}^{-1}_0)^{a\sigma}|y}=0\, ,
\label{matrel}
\en
where
\eq
\Delta(x-y)&=&\int\frac{d^dk}{(2\pi)^d}\frac{e^{-ik(x-y)}}{-k^2},\nn\\[2mm]
\Delta^{\rho\sigma}(x-y)&=&(a-1)\int\frac{d^dk}{(2\pi)^d}\frac{k^\rho k^\sigma}{k^4}e^{-ik(x-y)}\, .
\label{matrel1}
\en
The matrix elements of the operator $\omega$ are given by
\eq
\tr{y|\omega^{AB}|x}&=&-2Y^{AB}_\mu(y)
\frac{\partial}{\partial y_\mu}\delta(x-y)+c^{AB}(y)\delta(x-y),
\nn\\[2mm]
c(x)&=&-\partial_\mu Y_\mu-Y_\mu Y_\mu+\Lambda\, .
\en
The components of the above matrix can be written in the following form
\eq
\label{delta}
\tr{y|\omega^{ab}|x}&=&-2\Gamma^{ab}_\mu(y)\frac{\partial}{\partial y_\mu}\,\delta(x-y)
+b^{ab}(y)\delta(x-y),\nn\\[2mm]
\tr{y|\omega^{a\sigma}|x}&=&-2X^a(y)\frac{\partial}{\partial y_\sigma}\,\delta(x-y)
+b^{a\sigma}(y)\delta(x-y),\nn\\[2mm]
\tr{y|\omega^{\rho b}|x}&=&-2X^b(x)\frac{\partial}{\partial x_\rho}\delta(x-y)
+b^{\rho b}(x)\delta(x-y),\nn\\[2mm]
\tr{y|\omega^{\rho\sigma}|x}&=&\delta^{\rho\sigma}b(y)\,\delta(x-y),
\en
where
\eq
\label{delta1}
b^{ab}&=&-\partial_\mu\Gamma_\mu^{ab}-\Gamma_\mu^{ac}\Gamma_\mu^{cb}+4X^aX^b+\sigma^{ab}\, ,\nn\\[2mm]
b^{a\sigma}=b^{\sigma a}&=&
-\partial_\sigma X^a-\Gamma_\sigma^{ac} X^c
+\frac{1}{2}\,\gamma^{a\sigma}\, ,\nn\\[2mm]
b&=&X^aX^a+\rho\, .
\en
From Eq.~(\ref{Zexp}) after rather voluminous calculations one obtains
\eq\label{Zdet}
\frac{1}{2}\ln \det (D_0+\omega)&\!\!=\!\!&\frac{1}{2}\ln \det (D_0+\omega)\biggr|_{a=1}
+\lambda\int dx\biggl(\frac{1}{2}\,S_1-\frac{1}{4}\,S_2+\frac{1}{6}\,S_3-\frac{1}{8}\,S_4\biggr)
\nn\\[2mm]
&\!\!+\!\!&\mbox{UV-finite part}\, ,
\en
where the first term is given by Eq.~(\ref{Zmin}), and
\eq\label{S1234}
S_1&\!=\!&0\, ,
\nn\\[2mm]
S_2&\!=\!&(a-1)\biggl[-8\partial_\mu X^a b^a_\mu-b^a_\mu b^a_\mu\biggr]
-(2(a-1)+(a-1)^2)\biggl[2b^2\biggr]\, ,
\nn\\[2mm]
S_3&\!=\!&(a-1)\biggl[-24X^ab^{ab}X^b-12X^a\Gamma_\mu^{ab}b_\mu^b\biggr]
-(2(a-1)+(a-1)^2)\biggl[24bX^aX^a\biggr]\, ,
\nn\\[2mm]
S_4&\!=\!&(a-1)\biggl[32X^a\Gamma_\mu^{ac}\Gamma_\mu^{cb}X^b\biggr]
-(2(a-1)+(a-1)^2)\biggl[64X^aX^a X^bX^b\biggr]\, .
\en
Note that $S_1,\ldots ,S_4$ correspond to the loop with one, two, three, four 
external legs, see Eq.~(\ref{Zexp}). The quantity $S_1$ vanishes in the dimensional regularization, because it contains a no-scale integral.

Putting things together and using Eq.~(\ref{delta1}), we obtain
\eq
S&=&\frac{1}{2}\,S_1-\frac{1}{4}\,S_2+\frac{1}{6}\,S_3-\frac{1}{8}\,S_4
\nn\\[2mm]
&=&(a-1)\biggl[2b_\mu^a(\partial_\mu X^a+\Gamma_\mu^{ab}X^b)
+\frac{1}{4}\,b_\mu^ab_\mu^a
-16X^aX^aX^bX^b-4X^aX^b\sigma^{ab}\biggr]
\nn\\[2mm]
&+&\frac{1}{2}\,(2(a-1)+(a-1)^2)\biggl[\rho-3X^aX^a\biggr]^2\, .
\label{divZ}
\en
This is our final expression for the one-loop determinant in an arbitrary covariant gauge.

\section{Renormalization}
\label{sec:LECs}

At the next step, we continue Eq.~(\ref{divZ}) back to Minkowski space and
substitute explicit expressions, given in Eq.~(\ref{YLambda}).
Carrying out the summation over flavor indices, 
in the three-flavor case one obtains:
\eq\label{begin}
{\rm div}\,Z_{\rm one\;loop}&=&{\rm div}\,Z^{a=1}_{\rm one\;loop}
+\lambda(a-1)\frac{F_0^2}{2}\,\int dx\biggl\{
\langle [H_L,\Delta_\mu]^2\rangle
-\langle [H_R,\Delta_\mu]^2\rangle
\nn\\[2mm]
&+&\langle\sigma H_L^2\rangle
-\frac{ZF_0^2}{2}\,\langle[H_R,H_L]^2\rangle
-2\langle[H_R,\Delta_\mu]G^\mu\rangle-\frac{3}{4}\,\langle G_\mu G^\mu\rangle\biggr\}\, ,\nn\\
\en
where we have used the relation
\eq
D_\mu H_L=[H_R,\Delta_\mu]+G_\mu\, ,\quad\quad
G_\mu=u^\dagger c_{R\mu} Qu-uc_{L\mu}Qu^\dagger\, .
\en
It should be pointed out that the terms with $(a-1)^2$ have completely
cancelled in the final expression.

By using the equations of motion,
the equation (\ref{begin}) can be simplified to
\eq
{\rm div}\,Z_{\rm one\;loop}&=&{\rm div}\,Z^{a=1}_{\rm one\;loop}
+\lambda(a-1) F^2_0\int dx \biggr\{\nn\\
&&\frac{1}{4}\tr{(\chi \bar U^\dagger + \bar U\chi^\dagger + \chi^\dagger \bar U+\bar U^\dagger\chi )Q^2}\nn\\
&-&\frac{1}{4}\tr{(\chi \bar U^\dagger + \bar U\chi^\dagger )Q\bar U Q\bar U^\dagger+(\chi^\dagger \bar U+\bar U^\dagger\chi )Q \bar U^\dagger Q \bar U}\nn\\
&+&\frac{1}{4}\tr{(\chi \bar U^\dagger - \bar U\chi^\dagger )Q\bar U Q\bar U^\dagger
+(\chi^\dagger \bar U-\bar U^\dagger\chi )Q \bar U^\dagger Q \bar U}\nn\\
&+&\frac{1}{2}\tr{d_{\mu}\bar U^\dagger [c_{ R}^{\mu}Q,Q]\bar U
+d_{\mu}\bar U[c_{ L}^{\mu}Q,Q]\bar U^\dagger}+\frac{3}{4}\tr{c_{ R}^{\mu}Q
  \bar U c_{ L \mu}Q \bar U^\dagger }\nn\\
&-&\frac{3}{8}\tr{c_{ R}^\mu Q c_{ R \mu} Q + c_{ L}^\mu Q c_{ L\mu}Q}\biggr\}.
\label{fdivZ}
\en
The UV divergences in the electromagnetic LECs can be directly read off from
the expression above. Using the result of Ref.~\cite{Urech}, obtained for
$a=1$, we get (see also Eq.~(A.7) from Ref.~\cite{Knecht-Urech}):
\eq\label{eq:Sigma3}
\begin{array}{l l l}
\Sigma_1=\frac{3}{4},&
\Sigma_2=Z,&
\Sigma_3=-\frac{3}{4},\\[2mm]
\Sigma_4=2Z,&
\Sigma_5=-\frac{9}{4},&
\Sigma_6=\frac{3}{2}\,Z,\\[2mm]
\Sigma_7=0,&
\Sigma_8=Z,&
\Sigma_9=-\frac{1}{4}+\frac{1}{4}\,(1-a),\\[2mm]
\Sigma_{10}=\frac{1}{4}+\frac{3}{2}\,Z-\frac{1}{4}\,(1-a),&
\Sigma_{11}=\frac{1}{8}+\frac{1}{4}\,(1-a),&
\Sigma_{12}=\frac{1}{4}+\frac{1}{2}\,(1-a),\\[2mm]
\Sigma_{13}=\frac{3}{4}\,(1-a),&
\Sigma_{14}=-\frac{3}{8}\,(1-a),&
\Sigma_{15}=\frac{3}{2}+3Z+14Z^2,\\[2mm]
\Sigma_{16}=-3-\frac{3}{2}\,Z-Z^2,&
\Sigma_{17}=\frac{3}{2}-\frac{3}{2}\,Z+5Z^2.&\\
\end{array}\nn\\
\en
In case of two flavors, the first term of Eq.~(\ref{fdivZ}) can be further
simplified by using the following identity
\eq
&&\tr{(\chi \bar U^\dagger + \bar U\chi^\dagger + \chi^\dagger \bar U+\bar U^\dagger\chi
  )Q^2}\nn\\[2mm]
&-&\tr{(\chi \bar U^\dagger + \bar U\chi^\dagger )Q\bar U Q\bar U^\dagger+(\chi^\dagger \bar U+\bar U^\dagger\chi )Q \bar U^\dagger Q \bar U}
\nn\\[2mm]
&=&-\tr{\chi^\dagger \bar U+\bar U^\dagger\chi}\tr{Q^2}+ \tr{\chi^\dagger
  \bar U+\bar U^\dagger\chi}\tr{Q\bar U Q\bar U^\dagger}
\en
Using the effective Lagrangian, given in
Ref.~\cite{Knecht-Urech}, one reads off the divergent parts of the LECs in the  
two-flavor case as well
\eq
k_i=\lambda\sigma_i+k^r_i(\mu)\, ,\quad\quad i=1,\cdots 14\, .
\en
\eq\label{eq:Sigma2}
\hspace*{-1.cm}&&\hspace*{-.9cm}\begin{array}{l l l}
\sigma_1=-\frac{27}{3}-\frac{1}{5}\,Z, &
\sigma_2=2Z, &
\sigma_3=-\frac{3}{4},\\[2mm]
\sigma_4=2Z,&
\sigma_5=-\frac{1}{4}-\frac{1}{5}\,Z+\frac{1}{4}\,(1-a),&
\sigma_6=\frac{1}{4}+2Z-\frac{1}{4}\,(1-a),\\[2mm]
\sigma_7=0,&
\sigma_8=\frac{1}{8}-Z+\frac{1}{4}\,(1-a),&
\sigma_9=\frac{1}{4}+\frac{1}{2}\,(1-a),\\[2mm]
\sigma_{10}=\frac{3}{4}\,(1-a),&
\sigma_{11}=-\frac{3}{8}\,(1-a),&
\sigma_{12}=\frac{3}{2}-\frac{12}{5}\,Z+\frac{84}{25}\,Z^2,\\[2mm]
\sigma_{13}=-3-\frac{3}{5}\,Z-\frac{12}{5}\,Z^2,&
\sigma_{14}=\frac{3}{2}+3Z+12Z^2.&\\
\end{array}\nn\\
&&
\en
The equations (\ref{eq:Sigma3}) and (\ref{eq:Sigma2}) represent the main
result of the present paper. Note also that, through the renormalization group equations, these equations define the gauge dependence of the scale-dependent part of the LECs $K_i^r(\mu)$ and $k_i^r(\mu)$, respectively.

\section{Comparison to existing calculations}
\label{sec:comparison}

The dependence of some of the LECs on the gauge parameter has been studied
in Refs.~\cite{Moussallam,Anant,Scimemi}. We have explicitly checked that, in all
cases, our results agree with those from Refs.~\cite{Moussallam,Anant,Scimemi}.

Furthermore, as a useful check, we have verified that the scale-dependent part
of the combinations of the renormalized electromagnetic LECs, which appear in
the expressions for the pion, kaon and $\eta$-meson masses in 3-flavor ChPT~\cite{Urech},
\eq
C_1&=&6K_{1}^{r}+6K_{2}^{r}+5K_{5}^{r}+5K_{6}^{r}-6K_{7}^{r}-15K_{8}^{r}
-5K_{9}^{r}-23K_{10}^{r}-18K_{11}^{r},\nn\\[2mm]
C_2&=&K^r_8,\nn\\[2mm]
C_3&=&12K_{1}^{r}+12K_{2}^{r}-18K_{3}^{r}+9K_{4}^{r}+10K_{5}^{r}
+10K_{6}^{r}-12K_{7}^{r}-12K_{8}^{r}\nn\\[2mm]
&-&10K_{9}^{r}-10K_{10}^{r},\nn\\[2mm]
C_4&=&3K_{8}^{r}+K_{9}^{r}+K_{10}^{r},\nn\\[2mm]
C_5&=&6K_{1}^{r}+6K_{2}^{r}+5K_{5}^{r}+5K_{6}^{r}-6K_{7}^{r}-24K_{8}^{r}
-2K_{9}^{r}-20K_{10}^{r}-18K_{11}^{r},\nn\\[2mm]
C_6&=&3K_{1}^{r}+3K_{2}^{r}+K_{5}^{r}+K_{6}^{r}-3K_{7}^{r}-3K_{8}^{r}
-K_{9}^{r}-K_{10}^{r},\nn\\[2mm]
C_7&=&12K_{1}^{r}+12K_{2}^{r}-6K_{3}^{r}+3K_{4}^{r}+6K_{5}^{r}+6K_{6}^{r}-
12K_{7}^{r}-12K_{8}^{r}\nn\\[2mm]
&-&4K_{9}^{r}-4K_{10}^{r}\, ,
\en
does not depend on the gauge parameter $a$. The same statement holds for the
combinations of the LECs, which appear in the $\pi K$ scattering amplitude~\cite{Nehme}. Below, for illustration, we list some of these combinations: 
\eq
C_8&=&9(M^2_\pi+2M^2_K)K^r_8-M^2_\pi K^r_9+(17M^2_\pi+18M^2_K)K^r_{10}      +18(M^2_\pi +M^2_K)K^r_{11},\nn\\[2mm]
C_{9}&=&K_{5}^{r}+K_{6}^{r}+12K_{8}^{r}-6K_{10}^{r}-6K_{11}^{r},\nn\\[2mm]
C_{10}&=&18K_{3}^{r}-9K_{4}^{r}-12K_{8}^{r}+2K_{9}^{r}-34K_{10}^{r}-36K_{11}^{r}.
\en
Similar checks have been carried out in case of the 2-flavor ChPT, considering
the combinations of LECs, appearing in the pion masses and the $\pi\pi$
scattering amplitudes~\cite{Knecht-Urech,MMS}.

In Ref.~\cite{Haefeli}, the matching of the 2- and 3-flavor electromagnetic
LECs has been carried out. For illustration, consider one of the relations
from Ref.~\cite{Haefeli}:
\eq
k_5^r=\fc{6}{5}K_7^r+\fc{1}{5}K_8^r+\fc{4}{9}K_9^r-\fc{1}{5}K_{10}^r-\fc{1}{10}Z
\fc{1}{32{\pi^2}}(\ln{\fc{M_K^2}{\mu^2}+1})\, ,
\en
where $M_K$ stands for the kaon mass (at this order in $e$, there is no
difference between charged and neutral kaon masses). Note that only the loops with the
particles containing $s$-quark(s) (kaons and $\eta$) contribute to the matching
conditions. Consequently, since the photon loop is absent, the matching
condition should not contain the gauge parameter $a$ and be, therefore, gauge
invariant.
Inserting Eqs.~(\ref{eq:Sigma3}) and (\ref{eq:Sigma2}), 
we have checked that this is indeed the case for the scale-dependent part of
all relations listed in Ref.~\cite{Haefeli}. 

Finally, we wish to comment on the result of Ref.~\cite{GK}
(the divergent coefficients in 4 dimensions are
given in Ref.~\cite{Anant_review}, which 
contains a compilation of the earlier results). It turns out that the formulae
displayed there are not well suited for the direct comparison with 
our expressions. However,
it should be still pointed out that the divergent coefficients from table 2
of Ref.~\cite{Anant_review} 
contain logarithmic dependence on the gauge parameter.
It is clear that such a dependence can never arise in our framework. 
Note also that such a logarithmic dependence on the gauge parameter
arises from the photon mass term in the propagator and
is thus of the infrared origin.
Its appearance in the $\beta$-functions looks
counterintuitive to us.
Further, the logarithmic contributions in Ref.~\cite{Moussallam}, which were
mentioned in Ref.~\cite{Anant_review}, arise in various correlators and
not in the LECs.  
Consequently, these (scale independent)
contributions can not be identified with those from table 2 of 
Ref.~\cite{Anant_review}\footnote{We thank
  B.~Ananthanarayan and B. Moussallam for the discussions on this issue.}.

\section{Summary}
\label{sec:concl}

In this paper, we calculate the divergent part of the one-loop effective
action in ChPT with virtual photons in an arbitrary covariant gauge. Except
in the Feynman gauge, the differential operator in the action is of a 
non-minimal type, for which the standard technique, based on the heat kernel
expansion, is not directly applicable. Instead, we have resorted to a
straightforward perturbative expansion of the determinant. The final result
for the divergent part of the effective action is given in Eqs.~(\ref{Zdet}),
(\ref{S1234}) and (\ref{divZ}).
The divergent parts of the LECs in the 3- and 2-flavor ChPT are listed in
Eqs.~(\ref{eq:Sigma3}) and (\ref{eq:Sigma2}), respectively.

\bigskip

{\em Acknowledgments:} 
The authors thank J. Gasser for suggesting to investigate this problem 
and valuable comments. We would like to
thank B.~Ananthanarayan for current interest into the work and helpful
discussions. We thank B. Kubis, U.-G. Mei{\ss}ner and B. Moussallam
 for the interesting discussions.
This work is partly supported by the EU
Integrated Infrastructure Initiative HadronPhysics3 Project  under Grant
Agreement no. 283286. We also acknowledge the support by DFG (CRC 16,
``Subnuclear Structure of Matter'' and CRC 110, ``Symmetries and the Emergence
of Structure in QCD'') and by
the Shota Rustaveli National Science Foundation
(Project DI/13/02).
This research is supported in part by Volkswagenstiftung
under contract no. 86260.

\end{document}